%% file: main.tex
\documentclass[fleqn,usenatbib]{mnras}

\usepackage{newtxtext,newtxmath}
\usepackage[T1]{fontenc}

\DeclareRobustCommand{\VAN}[3]{#2}
\let\VANthebibliography\thebibliography
\def\thebibliography{\DeclareRobustCommand{\VAN}[3]{##3}\VANthebibliography}

\usepackage{graphicx}	% Including figure files
\usepackage{amsmath}	% Advanced maths commands
\usepackage{xcolor}
\usepackage{tikz-cd}
\usepackage{multicol}
\usepackage{url}

\newcommand{\eq}[1]{Equation~\ref{eq:#1}}
\newcommand{\fig}[1]{Figure~\ref{fig:#1}}
\renewcommand{\sec}[1]{Section~\ref{sec:#1}}

\title[Efficient Emulation of Large Volumes]{Efficiently emulating distribution functions in gigaparsec volumes for varying cosmological parameters}

\author[C. C. Lovell et al.]{Christopher C. Lovell,$^{1,2}$\thanks{E-mail: chris.lovell.astro@gmail.com (CCL)}
Max E. Lee,$^{3}$
William J. Roper,$^{4}$
Daniel Anglés-Alcázar,$^{5}$
Shy Genel,$^{6,7}$ \newauthor
Shivam Pandey$^{8}$
and Francisco Villaescusa-Navarro$^{9,7}$
\\
$^{1}$Kavli Institute for Cosmology, University of Cambridge, Madingley Road, Cambridge CB3 0HA, UK\\
$^{2}$Institute of Astronomy, University of Cambridge, Madingley Road, Cambridge CB3 0HA, UK\\
$^{3}$Department of Astronomy, Columbia University, MC 5246, 538 West 120th Street, New York, NY 10027, USA\\
$^{4}$Astronomy Centre, University of Sussex, Falmer, Brighton BN1 9QH, UK\\
$^{5}$Department of Physics, University of Connecticut, 196 Auditorium Road, U-3046, Storrs, CT, 06269, USA\\
$^{6}$Columbia Astrophysics Laboratory, Columbia University, 550 West 120th Street, New York, NY 10027, US\\
$^{7}$Center for Computational Astrophysics, Flatiron Institute, 162 5th Avenue, New York, NY 10010, USA\\
$^{8}$Department of Physics and Astronomy, Johns Hopkins University, Baltimore, MD 21218, USA\\
$^{9}$Department of Astrophysical Sciences, Princeton University, Peyton Hall, Princeton NJ 08544, USA\\
}

\date{Accepted XXX. Received YYY; in original form ZZZ}

\pubyear{\the\year{}}

\begin{document}
\label{firstpage}
\pagerange{\pageref{firstpage}--\pageref{lastpage}}
\maketitle

\begin{abstract}
We present a new method for emulating the halo mass function (HMF) and other distribution functions in large effective volumes, down to low halo masses, whilst simultaneously modifying large ranges of parameters, for a fraction of the cost of traditional periodic cosmological simulations.
We demonstrate the method by selecting small regions, $V \sim (50 \,h^{-1}{\rm Mpc})^3$, with a range of overdensities from the \textsc{Quijote} suite, consisting of tens of thousands of $(1 \,h^{-1}{\rm Gpc})^3$ $N$-body simulation volumes run with varying $\Lambda$CDM parameters.
We train a differentiable emulator, conditioned on the overdensity of the region and these global parameters, to reproduce the halo mass function in these regions.
We then successfully recover the global distribution of halo masses of the entire box by integrating over the overdensity distribution.
Our approach uses just $\sim\,$0.026\% of the original simulation volume, and suggests that suites of targeted `zoom' simulations, extracted from low resolution parent volumes, can be used to emulate large volume simulations at a fraction of the computational cost, whilst simultaneously pushing the dynamic range to much lower masses than can be achieved in periodic simulations.
We discuss emulation of other key dark matter and baryonic distribution functions, as well as higher order statistics, with implications for the interpretation of upcoming wide field surveys on observatories such as \textit{Euclid}, \textit{Roman} and \textit{Rubin}.
\end{abstract}


%%%%%%%%%%%%%%%%%%%%%%%%%%%%%%%%%%%%%%%%%%%%%%%%%%

%%%%%%%%%%%%%%%%% BODY OF PAPER %%%%%%%%%%%%%%%%%%

\input{intro}

\input{simulations}

\input{methods}

\input{global}
\input{results}
\input{discussion}
\input{conc}

\section*{Acknowledgements}

With thanks to Will Handley, David Yallup, Benjamin Wandelt, Matthieu Schaller, Aswin Vijayan, Francisco Maion, Adrien Bayer and Abigail Lovell for valuable discussions.
CCL was supported by the research environment and infrastructure of the Handley Lab at the University of Cambridge.
MEL is supported by NSF grant DGE-2036197.
DAA acknowledges support from NSF CAREER award AST-2442788, an Alfred P. Sloan Research Fellowship, and Cottrell Scholar Award CS-CSA-2023-028 by the Research Corporation for Science Advancement.
The Flatiron Institute is supported by the Simons Foundation.
This work is supported by the Simons Collaboration on ``Learning the Universe''.
The authors used OpenAI GPT 5.2 to refine the code, and Claude Opus 4.6 was used to refine portions of the draft.
The authors take full responsibility for the final content.

\section*{Author Contributions}

We list here the roles and contributions of the authors according to the Contributor Roles Taxonomy (CRediT)\footnote{\url{https://credit.niso.org/}}.
\textbf{Christopher C. Lovell}: Conceptualization, Methodology, Investigation, Formal Analysis, Visualization, Software, Writing - original draft.
\textbf{Max E. Lee}: Methodology, Investigation, Writing - review \& editing.
\textbf{William J. Roper, Daniel Anglés-Alcázar, Shy Genel, Shivam Pandey}: Writing - review \& editing.
\textbf{Francisco Villaescusa Navarro}: Data curation.

%%%%%%%%%%%%%%%%%%%%%%%%%%%%%%%%%%%%%%%%%%%%%%%%%%
\section*{Data Availability}

The Quijote data is publicly available at \url{https://quijote-simulations.readthedocs.io}.
All codes used in the analysis are publicly available at \url{https://github.com/tessera-sim}. 
 
% The inclusion of a Data Availability Statement is a requirement for articles published in MNRAS. Data Availability Statements provide a standardised format for readers to understand the availability of data underlying the research results described in the article. The statement may refer to original data generated in the course of the study or to third-party data analysed in the article. The statement should describe and provide means of access, where possible, by linking to the data or providing the required accession numbers for the relevant databases or DOIs.



\bibliographystyle{mnras}
\bibliography{tessera,custom}

% \begin{thebibliography}{}
% \bibitem[]{}% Placeholder to allow compilation without running BibTeX.
% \end{thebibliography}

%%%%%%%%%%%%%%%%%%%%%%%%%%%%%%%%%%%%%%%%%%%%%%%%%%

%%%%%%%%%%%%%%%%% APPENDICES %%%%%%%%%%%%%%%%%%%%%

\appendix

\input{appendix}

%%%%%%%%%%%%%%%%%%%%%%%%%%%%%%%%%%%%%%%%%%%%%%%%%%

% Don't change these lines
\bsp	% typesetting comment
\label{lastpage}
\end{document}

%% file: intro.tex
\section{Introduction}\label{sec:intro}

In cold dark matter (CDM) cosmologies, virialised dark matter haloes are the fundamental unit of cosmic structure formation, connecting the large scale cosmic web down to galaxy and star formation scales. 
Seeded by tiny density fluctuations laid down during inflation \citep{bardeenSpontaneousCreationAlmost1983}, they merge hierarchically in a bottom-up fashion through mergers and accretion, eventually forming the largest collapsed objects in our Universe today, the massive galaxy clusters.
The seeding and growth of dark matter haloes is strongly dependent on the parameters of the underlying $\Lambda$CDM cosmology; the halo mass function (HMF), describing the comoving abundance of dark-matter haloes as a function of mass and redshift, is therefore a central ingredient in cosmology.
This is particularly the case at the high mass end, where cluster number-counts have long been known as a sensitive cosmological probe \citep[e.g.][]{ekeClusterEvolutionDiagnostic1996}.

Importantly, dark matter haloes are the sites within which visible galaxies form and reside \citep{wechslerConnectionGalaxiesTheir2018}, and it is through galaxies as tracers that we are able to learn about their host haloes, and the larger matter distribution that they in turn trace.
The next generation of cosmological surveys, on observatories such as Euclid and Rubin \citep{euclidcollaborationEuclidOverviewEuclid2025,ivezicLSSTScienceDrivers2019}, will measure the positions and properties of billions of galaxies, and tens of thousands of clusters, with unprecedented precision, promising to revolutionize our understanding of dark matter and dark energy.
These experiments typically employ phenomenological models to map the relationship between galaxies and dark matter haloes.
Such models require precision estimates for the underlying HMF, ideally as a function of the cosmological parameters.

Analytical predictions for the HMF have been presented since the 1970's, including the pioneering Press-Schechter and Extended Press-Schechter formalisms \citep[][]{pressFormationGalaxiesClusters1974,bondExcursionSetMass1991}, which utilised relative peak heights to present a `universal' form of the HMF across cosmology and redshift.
However, the most accurate predictions of the HMF come from empirical estimates using the outputs of cosmological $N$-body simulations.
These simulations model non-linear gravitational collapse at late times and small scales, and can also reveal deviations from universality that cannot be accounted for analytically.
A number of these simulations have been run at the fiducial cosmology of the time to predict the HMF \citep{jenkinsMassFunctionDark2001,warrenPrecisionDeterminationMass2006}, or on a limited set of cosmologies to build fitting functions \citep{tinkerHaloMassFunction2008,crocceSimulatingUniverseMICE2010,collaborationEuclidPreparationXXIV2023}.
However, as forthcoming surveys push statistical errors below the per cent level \citep{wuImpactTheoreticalUncertainties2010}, analytic fitting functions calibrated on a limited set of cosmologies can become a leading theoretical systematic \citep[see e.g.][]{bocquetHaloMassFunction2016,castroImpactBaryonsHalo2021}.

A complementary approach is to \emph{emulate} the HMF; run a designed suite of $N$-body simulations spanning a target parameter space, and train a surrogate model that can predict the HMF rapidly for new cosmologies \citep[e.g.][]{nishimichiDarkQuestFast2019,mcclintockAemulusProjectII2019,bocquetMiraTitanUniverseIII2020a}.
These emulators are now achieving per-cent-level accuracy \citep[e.g.][]{saez-casaresEMANTISEmulatorFast2024,ruanEmulatorbasedHaloModel2024,shenAemulus$n$Precision2025,chenCSSTCosmologicalEmulator2025}, utilising efficient space-filling sampling schemes such as latin hypercubes or Sobol sequences.
The latest models explore large parameter ranges, high halo masses in large comoving volumes, as well as different halo definitions.
These models can then be used in traditional likelihood-explicit inference schemes, as well as more modern simulation based inference (SBI) schemes, to directly infer posteriors on the cosmological parameters utilising these relatively expensive numerical simulations of high mass halo and cluster abundances \citep[e.g.][]{regameyGalaxyClusterCount2025,payerneSimulationbasedCosmologicalInference2026a}.

However, it is well known that baryons shape the profiles and integrated properties of haloes \citep[e.g.][]{gebhardtCosmologicalBackreactionBaryons2026a}.
Processes such as supernova-driven outflows, AGN feedback, and reionisation can redistribute or expel gas from haloes, reducing the central gravitational potential and causing dark matter to migrate outward, a process known as dark matter heating or cusp-core transformation.
At the high-mass end, AGN feedback is particularly effective at quenching star formation and driving large-scale outflows that alter halo profiles. 
These baryonic effects imprint themselves on the halo mass function, suppressing the abundance of haloes relative to dark-matter-only predictions, particularly at intermediate masses where stellar and AGN feedback are most efficient \citep[e.g.][]{sawalaAbundanceNotJust2013a,cuiEffectActiveGalactic2014}.
Hydrodynamic simulations, which self-consistently model the dark matter and baryonic components, can accurately capture these effects.
These models utilise subgrid prescriptions for sub-resolution physical processes that cannot be modelled from first principles.

However, running a sufficient number of hydrodynamic simulations, down to sufficient resolution to accurately model galaxy evolution processes, is a significant challenge.
The BAHAMAS and FLAMINGO projects \citep{mccarthyBahamasProjectCalibrated2017,schayeFLAMINGOProjectCosmological2023} explored the Gigaparsec regime, but at particle mass resolutions of between $10^{8 - 10} \; {\rm M_{\odot}}$.
Higher resolution hydrodynamic simulations, that can accurately resolve the knee of the galaxy stellar mass function and the behaviour of satellites, are typically run in volumes of $\sim$(100 $\rm Mpc$)$^3$ \citep[see][for a review]{vogelsbergerCosmologicalSimulationsGalaxy2020}.
The recent state-of-the-art \textsc{Colibre} simulations ran a single simulation at $10^6 \, {\rm M_{\odot}}$ resolution in a $(200 \, {\rm Mpc})^3$ box, one of the largest at this mass resolution run to date \citep{schayeCOLIBREProjectCosmological2025}.
Such volumes only contain a handful of lower mass ($\sim 10^{14}{\; \rm M_{\odot}}$) galaxy clusters, and do not simulate the high mass ($> 10^{14.5}{\; \rm M_{\odot}}$) clusters necessary for probing the exponential tail of the halo mass function.
Additionally, these models typically assume only a single fiducial cosmology, and do not vary the underlying astrophysical model.
The CAMELS simulations \citep{villaescusa-navarroCAMELSProjectCosmology2021,niCAMELSProjectExpanding2023}, a large suite of thousands of hydrodynamic simulations varying a range of cosmological and astrophysical parameters, have been run in $(25 \; h^{-1}{\rm Mpc})^3$ volumes to date.
These simulations enable field level emulation as a function of cosmological and astrophysical parameters \citep{villaescusa-navarroCAMELSMultifieldData2022,desantiRobustFieldlevelLikelihoodfree2023,lovellHierarchyNormalizingFlows2023,gebhardtCosmologicalBackreactionBaryons2026a} and simulation-based inference of cosmological parameters marginalizing over uncertainties in baryonic effects \citep{villaescusa-navarroCosmologyOneGalaxy2022,hassanHIFLOWGeneratingDiverse2022,lovellLearningUniverseCosmological2025,iyerHowDoesFeedback2025a}.
However, the CAMELS simulation volumes are clearly insufficient for comparing to wide field surveys, and do not contain any cluster mass haloes.

One approach for overcoming this volume-resolution-parameter space trade-off is to employ `zoom' simulations \citep[e.g.][]{leeZoomingCARPoolGPLane2024}.
These take a small area of interest from a large, low resolution parent (typically $N$-body) simulation, and resimulate with added physics and/or higher resolution \citep{katzHierarchicalGalaxyFormation1993,tormenStructureDynamicalEvolution1997,stopyraGenetICNewInitial2021}.
This allows for sampling of rare environments whilst still maintaining the high fidelity required to model low mass haloes, and optionally the galaxy formation and evolution within those haloes.
If a sufficient number of zooms are run for a range of overdensities, they can be combined to produce composite distribution functions that mimic a much larger box.
\cite{crainGalaxiesIntergalacticMediumInteraction2009} pioneered this approach, using a suite of just 9 spherical hydrodynamical simulations selected from the dark matter-only (DMO) Millennium simulation.
By appropriately weighting these regions by their relative abundance in the parent simulation they created composite distribution functions, such as the galaxy stellar mass function, equivalent to those that would be measured given a hydrodynamic simulation at similar resolution ($1.45 \times 10^6 \, h^{-1} {\rm M_{\odot}}$) run in the entire Millennium volume.
This was unfeasible at the time, and is still beyond current computational capabilities; the recent MillenniumTNG simulations \citep{hernandez-aguayoMillenniumTNGProjectHighprecision2022,pakmorMillenniumTNGProjectHydrodynamical2023} are run with a baryonic particle resolution of $2 \times 10^{7} \, h^{-1} {\rm M_{\odot}}$.

The FLARES simulations \citep{lovellFirstLightReionization2021,vijayanFirstLightReionization2021} further demonstrated this approach at high redshift (down to $z = 5$), by resimulating 40 regions with full hydrodynamics at $1.8 \times 10^6 \, {\rm M_{\odot}}$ resolution, selected from a $(3.2 \; {\rm Gpc})^3$ parent DMO simulation with a range of overdensities.
This significantly extended the dynamic range of the predicted distribution functions, including the brightest, rarest galaxies, allowing them to make predictions for wide field surveys in the Epoch of Reionization, such as Euclid and Roman.
% as well as the earliest galaxy populations at $z > 10$ \citep{wilkinsFirstLightReionization2023}.
Most recently, a number of authors have shown how machine learning approaches can leverage zoom simulations to include uncertainties in one point distributions, as well as emulate higher order statistics \citep{lovellMachineLearningApproach2022,maltzFirstLightReionization2025}.

In this work we take inspiration from these advances in emulation and zoom simulation approaches, and present a new framework predicting distribution functions for arbitrary parameters, trained on the Quijote simulations \citep{villaescusa-navarroQuijoteSimulations2020}.
Rather than using full periodic boxes to train our emulator, we select small regions from a suite of `parent' simulations, and train an emulator conditioned on the cosmology of the parent \textit{and} the selected regions overdensity.
Using this emulator, we recover the \textit{global} distribution function by integrating the emulator over the overdensity distribution, which is cheap to accurately obtain from low-resolution periodic simulations.
Using selected regions as proxies for true zoom simulations, we demonstrate the efficacy of this method for combining zoom simulations.
We use the halo mass function (HMF) as a challenging test distribution to replicate.

Our analysis is arranged as follows.
In \sec{simulations} we describe the Quijote simulation suite and the utilised data products.
In \sec{methods} we outline the general framework, and present our conditional emulator for the overdensity and cosmology conditioned HMF in \sec{norm_flow}.
In \sec{global} we describe how we obtain the global HMF distribution from this emulator, and present our results in \sec{results}.
Finally, we discuss applications of our approach to other distribution functions, higher order statistics, and within zoom simulation suites of dark-matter only and full hydrodynamic models in \sec{discussion}.
We state our conclusions in \sec{conc}.

%% file: simulations.tex
\section{The Quijote Simulation Suite}
\label{sec:simulations}

We use the publicly available \textsc{Quijote} simulation suite \citep{villaescusa-navarroQuijoteSimulations2020}, which provides a large set of gravity-only $N$-body simulations.
In particular, we make use of the BSQ suite \citep[Big Sobol Sequence;][]{bairagiBIGSOBOLSEQUENCE2025}, which consists of thousands of individual simulations.
Each simulation was run with \textsc{Gadget-III} \citep{springelCosmologicalSimulationCode2005} assuming a flat $\Lambda$CDM cosmology, and follows the evolution of $512^3$ dark-matter particles in a periodic comoving volume of $(1000\,h^{-1}\,\mathrm{Mpc})^3$, with a different random initial seed for each realization.
Initial conditions are generated at $z=127$ using second-order Lagrangian perturbation theory (2LPT)\footnote{Using the code described here: \url{https://cosmo.nyu.edu/roman/2LPT/}}, with input matter power spectrum and transfer functions obtained by rescaling the $z=0$ outputs from \textsc{CAMB} \citep{lewisEfficientComputationCosmic2000a}.
The gravitational softening length is set to 50 $h^{-1}\,\mathrm{kpc}$.

The BSQ suite consists of $32{,}768$ simulations varying five standard cosmological parameters ($\Omega_\mathrm{m}$, $\Omega_\mathrm{b}$, $h$, $n_s$, $\sigma_8$), where the cosmological parameters are arranged in a Sobol low-discrepancy sequence with bounds
\begin{align*}
  \Omega_\mathrm{m} &\in [0.10, 0.50] \\
  \Omega_\mathrm{b} &\in [0.02, 0.08] \\
  h &\in [0.50, 0.90] \\
  n_s &\in [0.80, 1.20] \\
  \sigma_8 &\in [0.60, 1.00]
\end{align*}
% in the LH and BSQ suites
The remaining cosmological parameters are fixed to $M_\nu=0\,\mathrm{eV}$, $w=-1$, and $\Omega_k=0$.
Haloes are identified using the Friends-of-Friends (FoF) algorithm \citep{davisEvolutionLargescaleStructure1985} with a linking length $b = 0.2$, and no additional unbinding step.
We train using all haloes containing greater than 20 particles, and use the total mass of the halo as our target variable\footnote{We acknowledge the impact of halo mass definition on cluster cosmology, but note that only total FOF halo masses are available for the BSQ suite.}.
This gives a minimum halo mass of $10^{13.3} \; {\rm M_{\odot}}/h$ for the highest value of $\Omega_m$, 
\begin{align}
  M_{\rm min} = 20 \times m_p = \frac{20 \,\Omega_m \, \rho_{\rm crit} \, L^3}{N_p} \;.
\end{align}
where $N_p$ is the overall number of particles in the simulation, $L$ is the periodic box side length, and $m_p$ is the particle mass.

We use the $z = 0$ halo catalogues and precomputed 3D density fields from each simulation.
The 3D fields are computed using a cloud in cell (CIC) scheme on a 256$^3$ grid.
This leads to a voxel volume of $(3.91 \; h^{-1} {\rm Mpc})^3$.
The overdensity in voxel $i$ is given by
\begin{align}
  \delta_i = \frac{\rho_i}{\bar{\rho}} - 1 ,
  \label{eq:overdensity}
\end{align}
where $\rho_i$ is the density in voxel $i$ and $\bar{\rho}$ is the mean particle density per voxel in that simulation.
We also measure the number density of haloes per unit volume per dex at mass bin $j$ within each voxel $s$, $n_{sj}$, by counting haloes in $J$ mass bins with edges $\log_{10}M_j$ and widths $\Delta \log_{10} M_j \equiv \log_{10} M_{j+1}-\log_{10} M_j$ and normalising by the volume of the voxel and the bin width.

\begin{figure}
	\includegraphics[width=\columnwidth]{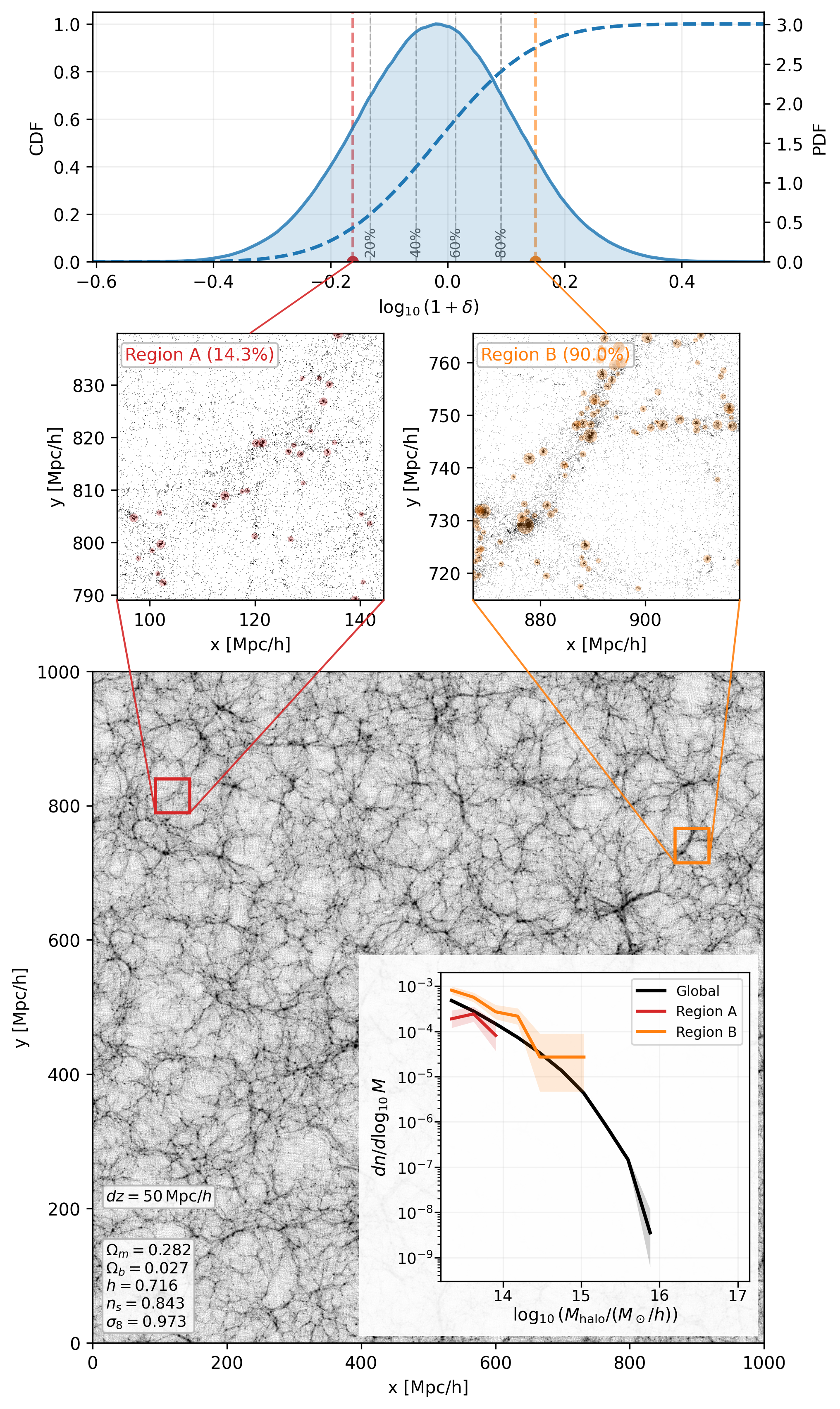}
    \caption{
      Figure demonstrating the region selection procedure. Bottom panel shows a density slice of one of the Quijote BSQ simulations, with depth $50 \, h^{-1}{\rm Mpc}$.
      The two inset panels show the density distribution in two selected regions, one with high overdensity and one low, each $(50 \, h^{-1}{\rm Mpc})^3$ in volume.
      Haloes in each region are shown with circles whose radius is proportional to their mass.
      The inset plot to the bottom right shows the halo mass function (HMF) in each region, as well as the HMF of the global parent simulation.
      The top panel shows the global overdensity distribution of regions of the same volume as a PDF, as well as the CDF of this same distribution, with each region's overdensity and respective ranked percentile shown.
    }
    \label{fig:illustration}
\end{figure}

We then combine voxels together into what we call \textit{regions}.
The overdensity $\delta_S$ and combined halo number density in bin $j$, $n_{Sj}$, of a region $S$ containing $\eta$ voxels is then the mean of its constituent voxel overdensities and halo number densities, respectively,
\begin{align}
  \delta_S = \frac{1}{\eta} \sum_{s \in \eta} \delta_s \, , \;\;\;\;\;\;\; &n_{Sj} = \frac{1}{\eta} \sum_{s \in \eta} n_{sj} \; .
\end{align}
Each mass bin $j$ in each region then contains $N_{Sj} = \Delta \log_{10} M_j n_{Sj} V_S$ haloes, where $V_S$ is the total volume of the region.
In the sections below we experiment with training using different region sizes by varying the number of voxels combined.
\fig{illustration} shows an example simulation, with two different regions selected.
Each region combines $13^3$ voxels together, for a total volume of $(50.78 \; h^{-1}{\rm Mpc})^3$.
It also shows the overall overdensity distribution, where these regions lie in overdensity and percentile space, and their individual region HMFs compared to the HMF of the whole simulation.

%% file: methods.tex
\section{Methods}
\label{sec:methods}

The general proposed framework is as follows.
Given a one-point distribution function $\Phi$ that depends on an independent variable $m$ (in this case the HMF, dependent on mass), as well as some global parameters $\theta$ (e.g. redshift, cosmological parameters) and the local overdensity of a region of characteristic size $R$, $\delta_R$, we can build a neural density estimator for this distribution function,
\begin{equation}
    \hat{\Phi}(m|\boldsymbol{\theta}, \delta_R, R) \sim
    \Phi(m|\boldsymbol{\theta}, \delta_R, R) + \epsilon(m|\boldsymbol{\theta}, \delta_R, R),
\end{equation}
where $\hat{\Phi}$ is our learnt emulator, and $\epsilon$ defines the bias of the emulator, which may be cosmology, overdensity or scale dependent.
The overdensity is that of the matter field smoothed with a filter $W$ of characteristic scale $R$,
\begin{equation}
    \delta_R(\boldsymbol{x}) = \int \delta(\boldsymbol{x}')\,
    W(|\boldsymbol{x}-\boldsymbol{x}'|; R)\, \mathrm{d}^3x' ,
\end{equation}
where in this work $W$ is a cubic top-hat of side $R = V_S^{1/3}$.
We can also model the distribution of overdensities, conditioned on the global parameters, $p(\delta|\boldsymbol{\theta})$.
The global distribution function can then be obtained by integrating the product of the overdensity PDF and the distribution function emulator over overdensity,
% \begin{equation}
%     \Phi_{\text{global}}(m|\boldsymbol{\theta}) = \int \Phi(m|\boldsymbol{\theta}, \delta) \, p(\delta|\boldsymbol{\theta}) \, d\delta \sim \int \hat{\Phi}(m|\boldsymbol{\theta}, \delta) \, p(\delta|\boldsymbol{\theta}) \, d\delta    
% \end{equation}
\begin{align}
    \Phi_{\rm global}(m|\boldsymbol{\theta})
    &= \int \Phi(m|\boldsymbol{\theta}, \delta_R, R)\, p(\delta_R|\boldsymbol{\theta}, R)\,
      \mathrm{d}\delta_R \\
    &\sim \int \hat{\Phi}(m|\boldsymbol{\theta}, \delta_R, R)\,
      p(\delta_R|\boldsymbol{\theta}, R)\, \mathrm{d}\delta_R .
\end{align}
As long as the emulator is precise and unbiased ($\epsilon \to 0$) across the entire range of overdensities for a given $\theta$, and the numerical integration is sufficiently accurate, this will return the global distribution function for arbitrary parameters $\theta$.
The left hand side carries no $R$ dependence: the smoothing scale is an internal choice of the framework, and the $R$ dependence of the conditional term must cancel against that of the overdensity PDF.
We assume here that the overdensity distribution $p(\delta_R | \boldsymbol{\theta}, R)$ is simple to obtain from a low resolution DMO simulation on sufficiently large scales $R$.\footnote{We have tested whether the overdensity distribution is converged on the scales used in this paper, and find that the percentiles do not change by more than 0.5\% on scales of $R = 27 \; {\rm Mpc}/h$, and even less on larger scales. This leads to negligible impact on the HMF when propagated through the emulator.}

In the rest of this section we describe the construction of our emulator $\hat{\Phi}(m|\boldsymbol{\theta}, \delta_R, R)$ for the HMF, and our integration approach for obtaining the global HMF.
For clarity we drop the $R$ subscript from $\delta$ and the explicit $R$ dependence in the rest of this paper.
We note that this scheme is generalisable to other one-point distribution functions and scaling relations.

\subsection{Conditional emulator for the HMF}
\label{sec:norm_flow}

\begin{figure*}
	\includegraphics[width=\textwidth]{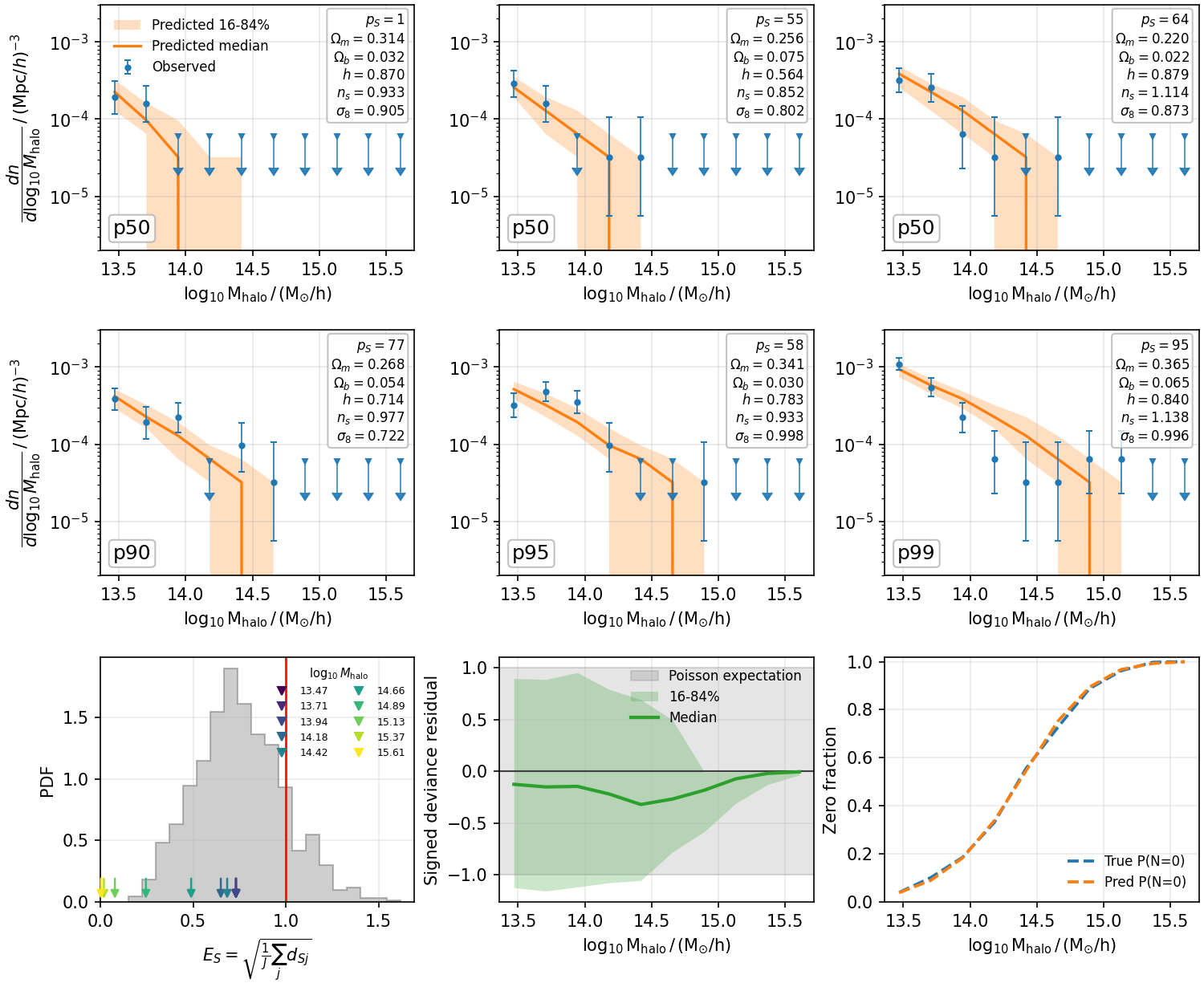}
    \caption{
        Evaluation of the region level conditional HMF emulator. 
        \textit{Top two rows}: six held-out examples selected at the 50th, 90th, 95th, and 99th percentiles of the Poisson deviance distribution, $E_S$, shown at the bottom left of each sub-panel. Each panel also show the global cosmological parameters, $\boldsymbol{\theta}$, and the overdensity percentile of the region, $p_S$. 
        \textit{Bottom-left}: distribution of $E_S$ across all test regions.
        \textit{Bottom-middle}: mass-dependent signed deviance residual shown as median and $16$--$84$ percentile bands over regions.
        \textit{Bottom-right}: calibration versus mass, including empirical 68\% and 95\% predictive coverage and zero-count calibration.
    }
    \label{fig:conditional_hmf_examples}
\end{figure*}

To aid training we decompose the HMF at a given mass into a global baseline and a residual, 
\begin{align}
    \log_{10} n_{Sj}(\boldsymbol{\theta},\delta) = \log_{10} n^{\mathrm{base}}_j + r_{Sj}(\boldsymbol{\theta},\delta). 
\end{align}
where the baseline in bin $j$ is independent of cosmology and overdensity.
We use a baseline that is derived from all selected \textit{training} regions, and is simply the aggregated mean HMF from these regions,\footnote{We also tested using the \cite{tinkerHaloMassFunction2008} mass function at some fiducial cosmology as a baseline, but found that this led to poorer performance compared to the average of all training regions.}
\begin{align}
    n^{\rm base}_{j} = \frac{1}{\Pi}\sum_{S=1}^\Pi n_{Sj} ,
\end{align}
where $\Pi$ is the total number of selected training regions.
Example `observed' HMFs from regions with different overdensities and from simulations with different cosmologies are shown in \fig{conditional_hmf_examples}.
In the majority of regions the highest mass bins are zero valued, which we indicate with arrows to show upper bounds on the number density.

We then train a conditional emulator to predict $r_{Sj}$.
We condition on both the overdensity \textit{percentile} of the region, $p_S$, the cosmological parameters, $\theta$, and the bin halo mass $m_j$; our feature vector is then $x_{Sj} = [p_S,\theta,m_j]$, and our predictor vector is $y_{Sj} = n_{Sj}$.
The overdensity percentile is measured with respect to the distribution of overdensities in the parent box from which the region is selected. 
We discuss in greater detail why we choose to condition on the overdensity percentile in \sec{global}.
We apply z-score normalisation to our features and predictors ($\tilde{x}_S$, $\tilde{y}_S$), where the mean and standard deviation are computed on the training set only.
% Each layer of the flow $\ell = 1, \ldots, L$ applies an affine transformation,
% \begin{align}
%   z^{(\ell)}_{Sj} = z^{(\ell-1)}_{Sj} \, \exp\!\big(s_\ell(\tilde{x}_{Sj})\big) 
%   + t_\ell(\tilde{x}_{Sj}),
% \end{align}
% where $z^{(0)}_{Sj} \sim \mathcal{N}(0,1)$ is the initial latent draw, and $s_\ell$ and $t_\ell$ are neural networks outputting per-layer scale and shift parameters. 
% The final output $r_{Sj} = z^{(L)}_{Sj}$ is the predicted residual.
% This architecture shares flow weights across mass bins while allowing bin-specific behavior via
% conditioning on \(m_j\).
We model the residual in each bin as a feature-dependent Gaussian,
\begin{equation}
r_{Sj}\,|\,\tilde{x}_{Sj} \sim
\mathcal{N}\!\left(\mu(\tilde{x}_{Sj}),\,\sigma^2(\tilde{x}_{Sj})\right) \;.
\end{equation}
In practice we draw $z^{(0)}_{Sj}\sim\mathcal{N}(0,1)$
and apply $L$ successive affine transformations,
\begin{equation}
z^{(\ell)}_{Sj} = z^{(\ell-1)}_{Sj}
\exp\!\left(s_\ell(\tilde{x}_{Sj})\right) + t_\ell(\tilde{x}_{Sj})\,,
\qquad r_{Sj} = z^{(L)}_{Sj}\,, 
\end{equation} 
where $s_\ell$ and $t_\ell$ are neural networks.
Since neither depends on the latent, the composition collapses to a single affine map.\footnote{We tested using a more expressive normalising flow model for the latent residual, but found similar results to our fiducial model. The normalising flow model is included with the source code, as it may be useful for modelling for more complex distribution functions in future work.}
% , with
% $\sigma = \exp\left(\sum_\ell s_\ell\right)$ and
% $\mu = \sum_\ell t_\ell \prod_{m>\ell} \exp(s_m)$.
%  the layered parameterisation affects optimisation rather than expressivity.

The predicted halo count $\tilde{N}_{Sj}$ in each region is modelled as a Poisson draw given the model-predicted rate $\hat{\lambda}_{Sj}$,
\begin{align}
\tilde{N}_{Sj} \mid \hat{\lambda}_{Sj} &\sim \mathrm{Poisson}(\hat{\lambda}_{Sj}), \\
\hat{\lambda}_{Sj} &= V_S\,\Delta\log_{10}M_j\,10^{(\log_{10} n^{\mathrm{base}}_j + r_{Sj})},
\end{align}
where $V_S$ is the region volume.
The Poisson--lognormal marginal likelihood has no closed form, so we approximate it by Monte Carlo, drawing $K$ independent residual samples per region and averaging the resulting likelihoods.
% Since $\hat{\lambda}_{Sj}$ depends on the latent draw $z^{(0)}_{Sj}$, the likelihood is intractable to marginalise analytically.
% We instead approximate it via Monte Carlo, drawing $K$ independent latent samples per region and averaging the resulting likelihoods.
The Poisson log-likelihood for region $S$ under the $k$-th flow sample is
\begin{align}
\ell_S^{(k)} &= \sum_{j=1}^J N_{Sj}\log\hat{\lambda}_{Sj}^{(k)} - \hat{\lambda}_{Sj}^{(k)} - \log(N_{Sj}!),
\end{align}
where $N_{Sj}$ is the true count in bin $j$, and  $\hat{\lambda}_{Sj}^{(k)}$ is the predicted rate from the $k$-th latent draw.
The training objective averages over samples and regions,
\begin{align}
\mathcal{L}(\theta)
&= -\frac{1}{\mathcal{S}}\sum_{S=1}^{\mathcal{S}}
\log\Bigg(\frac{1}{K}\sum_{k=1}^K
\exp\big[\ell_S^{(k)}\big]\Bigg)
+ \frac{\lambda_w}{2}\|\theta\|_2^2,
\end{align}
where $\lambda_w$ is a weight-decay coefficient.
At inference time, for a given feature vector $\tilde{x}_{Sj}$, the conditional HMF is predicted by averaging the flow output over $K$ latent draws,
\begin{align}
\hat{n}_{Sj} \approx \frac{1}{K}\sum_{k=1}^K
10^{(\log_{10} n^{\mathrm{base}}_j + r_{Sj}^{(k)})}.
\end{align}

%% file: global.tex
\subsection{Predicting the global conditional HMF}
\label{sec:global}

In \sec{norm_flow} we described how we built a region level emulator for the HMF conditioned on the \textit{percentile} of the overdensity of that region $S$, $p_S$, relative to the overall distribution of overdensities in the given parent simulation.
This is in contrast to conditioning on the numerical value of the overdensity, $\delta_S$.
The main motivation for this approach is that it allows us to avoid modelling the overdensity distribution $p(\delta)$, and its dependence on cosmological parameters, explicitly when recovering the global distribution function.
Instead, we just sample in percentile space using a Monte Carlo approach, described below, in order to recover the global distribution.
It also naturally bounds the input, avoiding out of distribution errors for overdensities at the extremes.
Below we describe our Monte Carlo approach for obtaining the global HMF given this percentile-conditioned region HMF emulator.

We create a realisation of the HMF in a given global simulation volume by sampling multiple regions to build up a similar global volume.
For each Monte Carlo realisation of the global simulation, we tile $\Pi_{\rm eff} = \lfloor V_{\rm global}/V_{\rm S}\rceil$ independent regions, where $V_{\rm S}$ is the comoving volume of a single region and $V_{\rm global} = L^3$ is the global simulation volume.
Each region $S$ is assigned a random density percentile $p_S\sim\mathcal{U}(0,100)$, and the emulator produces a predicted number density $\hat{n}_{Sj}$ in mass bin $j$ given this overdensity percentile, and the cosmological parameters of the parent box, $\boldsymbol{\theta}$. 
The expected and realised halo count in region $S$ is then
\begin{align}
  &\hat{\lambda}_{Sj}(\boldsymbol{\theta}, p_S) = \hat{n}_{Sj}(\boldsymbol{\theta}, p_S)\; V_{\rm S}\; \Delta\log_{10}M_j \\
  &\hat{N}_{Sj} (\boldsymbol{\theta}, p_S) \sim \mathrm{Poisson}(\hat{\lambda}_{Sj}(\boldsymbol{\theta}, p_S)) \,.
\end{align}
The predicted global count in each mass bin is obtained by summing over all regions, effectively integrating out the overdensity percentile dependence\footnote{This implicitly ignores any super-survey modes beyond the parent box size, as these do not exist in our training sample. It also ignores any super-sample covariance beyond the region size; we comment on this in \sec{global_hmf}.},
\begin{equation}
  \hat{N}_j^{\rm global} (\boldsymbol{\theta}) \approx \sum_{S=1}^{\Pi_{\rm eff}} \hat{N}_{Sj} (\boldsymbol{\theta}, p_S) \,,
\end{equation}
which is then converted back to a \textit{global} number density via
\begin{equation}
  \hat{n}_j^{\rm global} (\boldsymbol{\theta}) = \frac{\hat{N}_j^{\rm global} (\boldsymbol{\theta})}{V_{\rm global}\;\Delta\log_{10}M_j} \,.
\end{equation}
Repeating this procedure yields $\{\hat{n}_j^{\rm global}\}$, from which we report the median and 16th--84th percentile interval as the predicted global HMF and its uncertainty.

\subsection{Building and training the model}

We build our model in \textsc{JAX}, to leverage Just-In-Time (JIT) compilation for improved training and inference performance on GPU, and also to enable differentiability of the model through autodiff.
This allows the trained model to be used in gradient based downstream optimisation schemes, such as Blackjax \citep{cabezas2024blackjax}.

We begin by training a fiducial model using a region size of $13^3$ voxels, which corresponds to a volume of $(50.78 \; h^{-1}{\rm Mpc})^3$.
We use 4096 BSQ simulations, and select two regions from each one, for a total of 8192 regions.
In the examples in \sec{how_many} where we select multiple regions within each parent simulation, we ensure that training samples are split at the simulation level to prevent data leakage between spatially correlated voxels within the same simulation box.
Using a fixed random seed we shuffle and partition into training (80\%), validation (10\%) and test (10\%) subsets.

We implement two sampling schemes in overdensity percentile space.
In the first, we select each region randomly in percentile space.
In the second we sample regions in the 95$^{\rm th}$-100$^{\rm th}$ percentile range with a higher probability.
This provides a biased training sample that can improve the performance at the high mass end of the HMF.
In our fiducial model, we oversample by a factor of $1.5$; we find that this leads to improved performance. 
We then measure the HMF in each selected region using 14 mass bins between ${\rm log_{10}} \, M_{\rm halo} / ({\rm M_{\odot}} / h) = 13.35-17.15$; the lower limit corresponds to the minimum halo mass for a region with $\Omega_m = 0.5$.
Given these measured HMFs across all our selected training regions, we truncate the bin distribution to those bins that have at least one non-zero halo count across all training regions\footnote{Ensuring coverage across all halo mass bins would require a more sophisticated sampling procedure, which we will explore in future work}.
For our fiducial region volume, number of training regions, and percentile selection, this gives 11 bins, with the highest mass bin cut at $10^{16} \, {\rm M_{\odot}}/h$.

We use the Adam optimiser \citep{kingmaAdamMethodStochastic2014} with $\beta_1{=}0.9$,
$\beta_2{=}0.999$ and $\epsilon{=}10^{-8}$. 
Global gradient clipping with a default clipping threshold $c{=}1$ is applied before the parameter update.
We reduce the learning rate by a factor of 0.5 if the validation loss has not improved after 80 steps.
Validation loss is evaluated every step using the same NLL objective on the held-out validation set.
Early stopping halts training when the validation loss has not improved by more than $3\times10^{-5}$ for 300 consecutive validation checks.

Hyperparameters are tuned with Optuna \citep{akibaOptunaNextgenerationHyperparameter2019} using a tree-structured Parzen estimator (TPE) sampler.
Each trial trains a model from scratch with an independent random seed and a fixed realization split, then evaluates it on the held-out test set.
The target objective metric is the median absolute fractional error of the predicted global HMF across test realizations.
The search space is summarised in Table~\ref{tab:optuna}, as well as our best fit resulting parameters.
Optuna's median pruner terminates under-performing trials early: a trial is pruned if its validation loss exceeds the median of all completed trials at the same training step, with 5 startup trials exempted from pruning.
We perform our hyperparameter optimisation on our fiducial simulation and region selection counts (4096 and 2 respectively), and use the same architecture in \sec{how_many} for different numbers of regions; we acknowledge that these parameters may not be optimal for different size training sets.

%% file: results.tex
\section{Results}\label{sec:results}

\begin{table}
  \centering
  \caption{Optuna hyperparameter search space.  ``Log-$\mathcal{U}$''
  denotes a log-uniform distribution; ``Cat.''\ denotes a categorical
  choice.}
  \label{tab:optuna}
  \begin{tabular}{llll}
    \hline
    Parameter & Range / values & Sampling & Best fit \\
    \hline
    Flow layers          & 2--20 (step 2)
                                          & Integer            & 4 \\
    Hidden dim           & \{16, 32, 64, 128\} & Cat.               & 64 \\
    Hidden layers        & 1--3           & Integer            & 1 \\
    MC samples $K$       & \{2, 4, 8\}    & Cat.               & 4 \\
    Learning rate        & $[10^{-4},\; 5\times10^{-3}]$
                                          & Log-$\mathcal{U}$  & $1\times10^{-3}$ \\
    Weight decay         & $[10^{-6},\; 10^{-2}]$
                                          & Log-$\mathcal{U}$  & $2\times10^{-6}$ \\
    Grad-clip norm       & $[0.1,\; 10]$    & Log-$\mathcal{U}$  & 1.0 \\
    \hline
  \end{tabular}
\end{table}

\begin{figure}
	\includegraphics[width=\columnwidth]{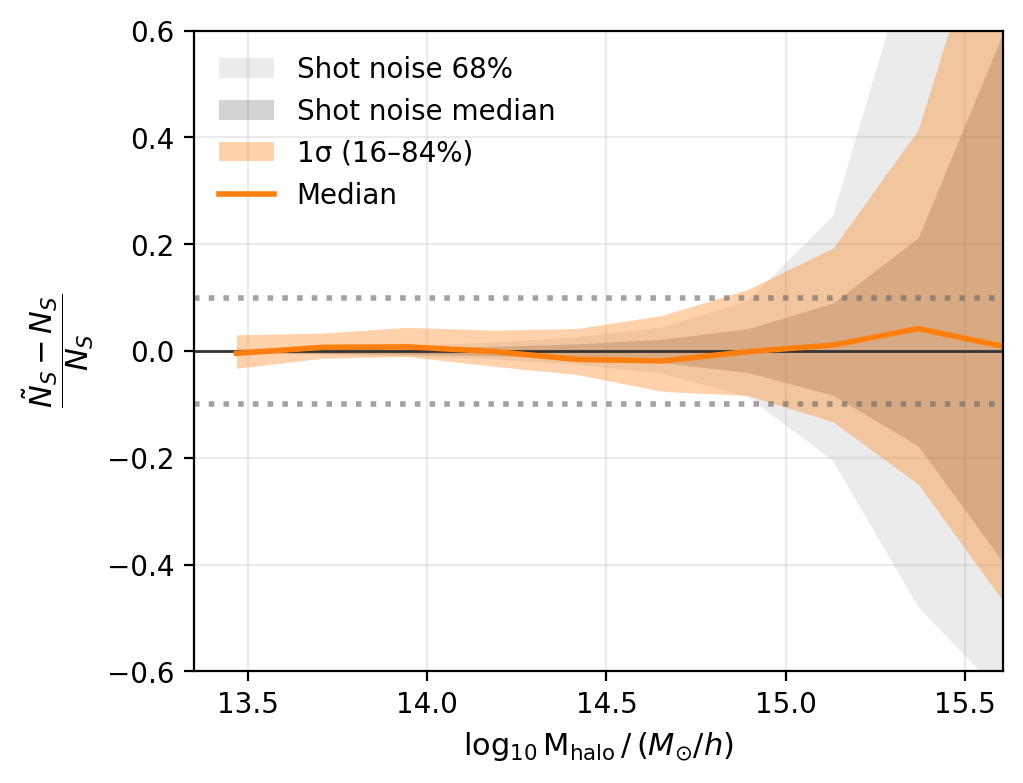}
    \caption{Fractional error in the \textit{global} HMF predictions across the test set. The Grey shaded regions show the median and 1$\sigma$ distribution of the shot noise across the test set. The solid orange line shows the median fractional error, and the shaded region the 1$\sigma$ distribution.
    }
    \label{fig:frac_error}
\end{figure}

\subsection{Conditional HMF region-level predictions}

In \fig{conditional_hmf_examples} we show six example predictions for the HMF in individual regions from the test set.
We split the examples by their overall error as represented by their deviance residual; the top row shows examples with median prediction errors, the second row examples in the 90$^{\rm th}$, 95$^{\rm th}$ and 99$^{\rm th}$ percentiles of the error distribution (across the test set).
The per-bin Poisson deviance is
\begin{align}
    d_{Sj}=2\!\left[\hat{\lambda}_{Sj}-N_{Sj}+N_{Sj}\ln\!\left(\frac{N_{Sj}}{\hat{\lambda}_{Sj}}\right)\right].
\end{align}
The region-level deviance is then given by the average over all bins,
\begin{align}
E_S=\sqrt{\frac{1}{J}\sum_{j=1}^{J} d_{Sj}}.
\end{align}
Each $d_{Sj}$ is a likelihood-ratio statistic, equal to $2\,\Delta\ln\mathcal{L}$ between the model prediction and a saturated model that perfectly reproduces the observed counts.
$\hat{\lambda}_{Sj}$ is strictly positive by construction, so $d_{Sj}$ is finite for all $N_{Sj}$.
$E_S \approx 1$ indicates that the residuals are consistent with Poisson scatter alone, and $E_S \gg 1$ indicates systematic model misfit beyond what Poisson noise can explain.
This is an approximate asymptotic calibration; 
for zero counts we adopt the standard convention $N \ln N \to 0$, giving $d_{Sj} = 2 \hat{\lambda}_{Sj}$, which vanishes as $\hat{\lambda}_{Sj} \to 0$; conversely a non-zero count in a bin with $\hat{\lambda}_{Sj} \to 0$ is penalised.
The former leads to $d_{Sj} \approx 0$, which pulls the mean of $E_S$ below unity.
The lower left panel of \fig{conditional_hmf_examples} shows the distribution of $E_S$ across all test regions.
The median lies close to but slightly below 1, which confirms that the residuals are consistent with Poisson scatter.
The coloured arrows indicate the median in different halo mass bins of $\sqrt{d_{Sj}}$; high mass bins, that are predominantly sparse, tend to bias the median below unity.
Importantly, all bins lie below 1, which indicates no systematic model misfit.

The bottom middle panel of \fig{conditional_hmf_examples} also shows the mass-dependent signed deviance residual,
\begin{align}
    \tau_{Sj} = \mathrm{sign}(N_{Sj} - \hat{\lambda}_{Sj})\,\sqrt{d_{Sj}},
\end{align}
plotted as the median and $16$--$84$th percentile bands over all test regions.
For an accurate Poisson model, the signed deviance residuals are approximately standard normal; a median consistent with zero indicates no systematic bias, while $16$--$84$th percentile bands within $\pm 1$ indicate that the residual scatter is consistent with Poisson noise.
There is a slight negative bias for bins with mass $\log_{10} \, M_{\rm halo} / ({\rm M_\odot} / h) \sim 14.5$, indicating that the emulator marginally over-predicts counts in these mass bins; we tested a model trained without oversampling the high-percentile bin, and found similar results.
The residual bands in low mass bins are consistent with Poisson noise, but narrow to zero at the high-mass end; this again reflects the dominance of zero-count bins where both the predicted and observed counts vanish, so the deviance carries no discriminating power.
The fit quality at these masses is better assessed by comparing the predicted zero-count probability $e^{-\hat{\lambda}_{Sj}}$ with the observed zero-count fraction across regions (bottom-right panel); the zero counts are predicted accurately in all mass bins.

Overall, the emulator accurately predicts the shape and normalisation of the HMF across the range of cosmologies, overdensity percentiles and mass bins in the test set, and effectively handles the transition to the zero-count regime at high masses.

\subsection{Global HMF predictions}
\label{sec:global_hmf}

We now assess the reconstruction of the global HMF using our conditional HMF model.
We first quantitatively assess the performance by showing the median and 1$\sigma$ distribution of the fractional error across the test set in \fig{frac_error}.
At higher masses shot noise contributes to the errors; we estimate the shot-noise uncertainty on the measured halo mass function using the Garwood interval \citep{garwoodFiducialLimitsPoisson1936} to properly model asymmetric confidence intervals in the small count regime.
We show the median and 1$\sigma$ distribution of the shot noise distribution, computed directly from the true counts, as the grey shaded areas.
The uncertainties are close to the shot noise contribution at high masses, and converge to a floor of around 5\% at lower masses.
The median is unbiased at low masses, and is biased to higher counts at higher masses, though this is deep within the shot noise regime.
This performance is competitive with other HMF emulators presented in the literature \citep[e.g.][]{mcclintockAemulusProjectII2019,bocquetMiraTitanUniverseIII2020a,saez-casaresEMANTISEmulatorFast2024,shenAemulus$n$Precision2025}.
Our global integration method implicitly ignores any covariance between independently sampled regions; this may lead us to underestimate the variance in our global predictions when summing over all regions, explaining the deficit compared to the whole-box shot noise error.
In future work we will explore methods for estimating this covariance term explicitly.

\begin{figure*}
	\includegraphics[width=0.98\textwidth]{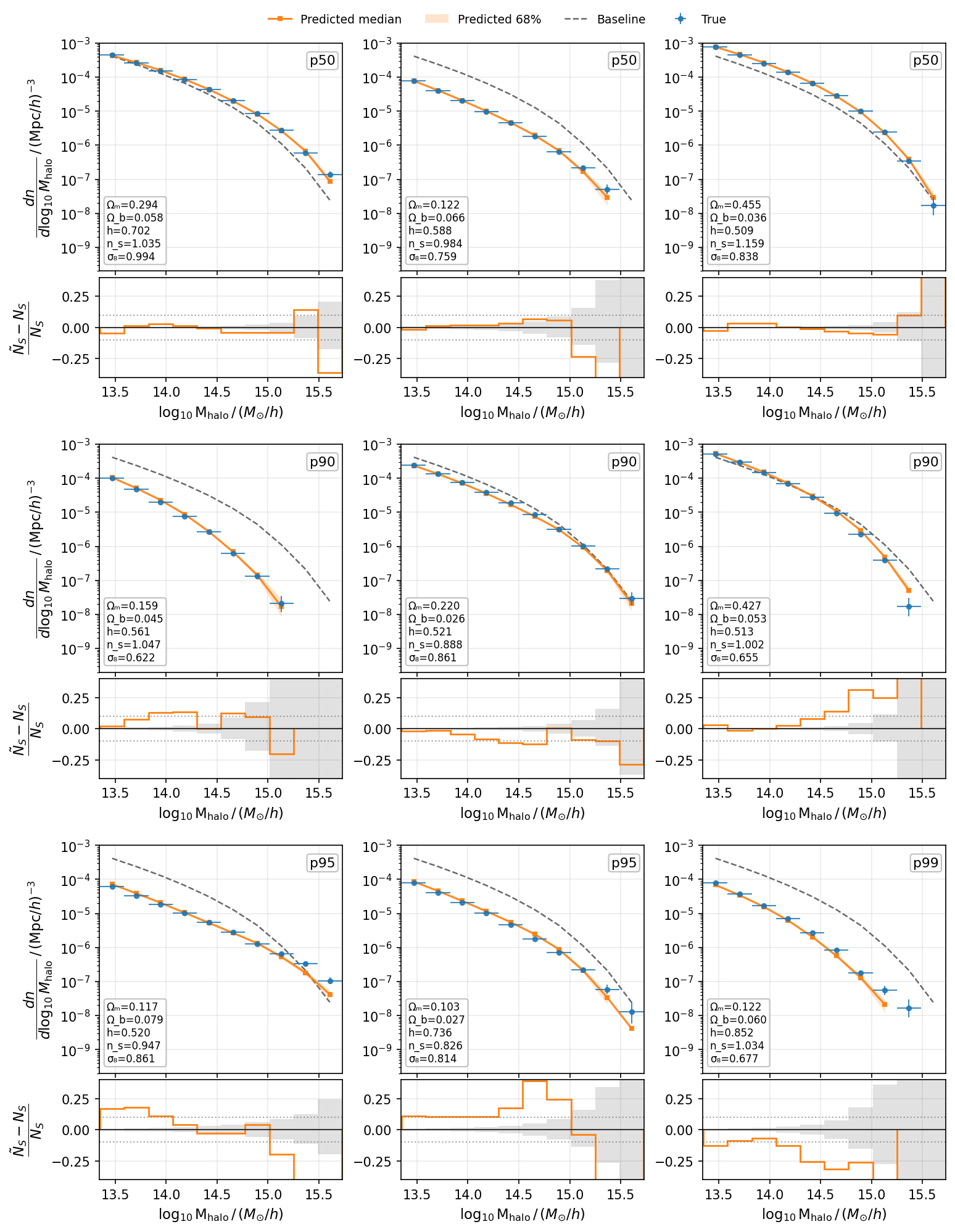}
    \caption{Example global HMF predictions from held out test simulations in the Quijote BSQ suite.
    Each panel shows the cosmological parameters of the held out simulation, the true HMF (blue) with Poisson confidence intervals, the baseline (calculated as the mean across all training regions, grey dashed), and the predicted HMF in orange with 68\% uncertainties.
    Each panel also shows the fractional error in the counts in each bin, alongside the estimated shot noise contribution to the uncertainties.
    We present examples close to the 50$^{\rm th}$, 90$^{\rm th}$, 95$^{\rm th}$ and 99$^{\rm th}$ percentiles of the median fractional error distribution.
    }
    \label{fig:global_hmf_examples}
\end{figure*}

\fig{global_hmf_examples} shows examples of the predicted global HMF using the fiducial model on held out test simulations from the Quijote BSQ suite, with examples ordered by the median fractional error in counts across all bins.
It also shows the input baseline measured from the training regions.
Overall, the method is able to accurately reproduce the shape and normalisation of the HMF across a wide range of cosmologies, with increasing fractional errors and uncertainty for higher mass bins as expected.
The shot noise contribution in terms of the fractional error in the counts is shown by the grey shaded region in each fractional error panel.
In most cases the uncertainty at the high mass end can be explained by this shot noise contribution, but in some cases the error exceeds this contribution.
For the majority of test regions the fractional errors are below 10\% in bins below $10^{15} \; {\rm M_{\odot}}$, and for 95\% of the test examples the errors are below 5\% in bins below $10^{14.5} \; {\rm M_{\odot}}$.

\begin{figure*}
	\includegraphics[width=1.0\textwidth]{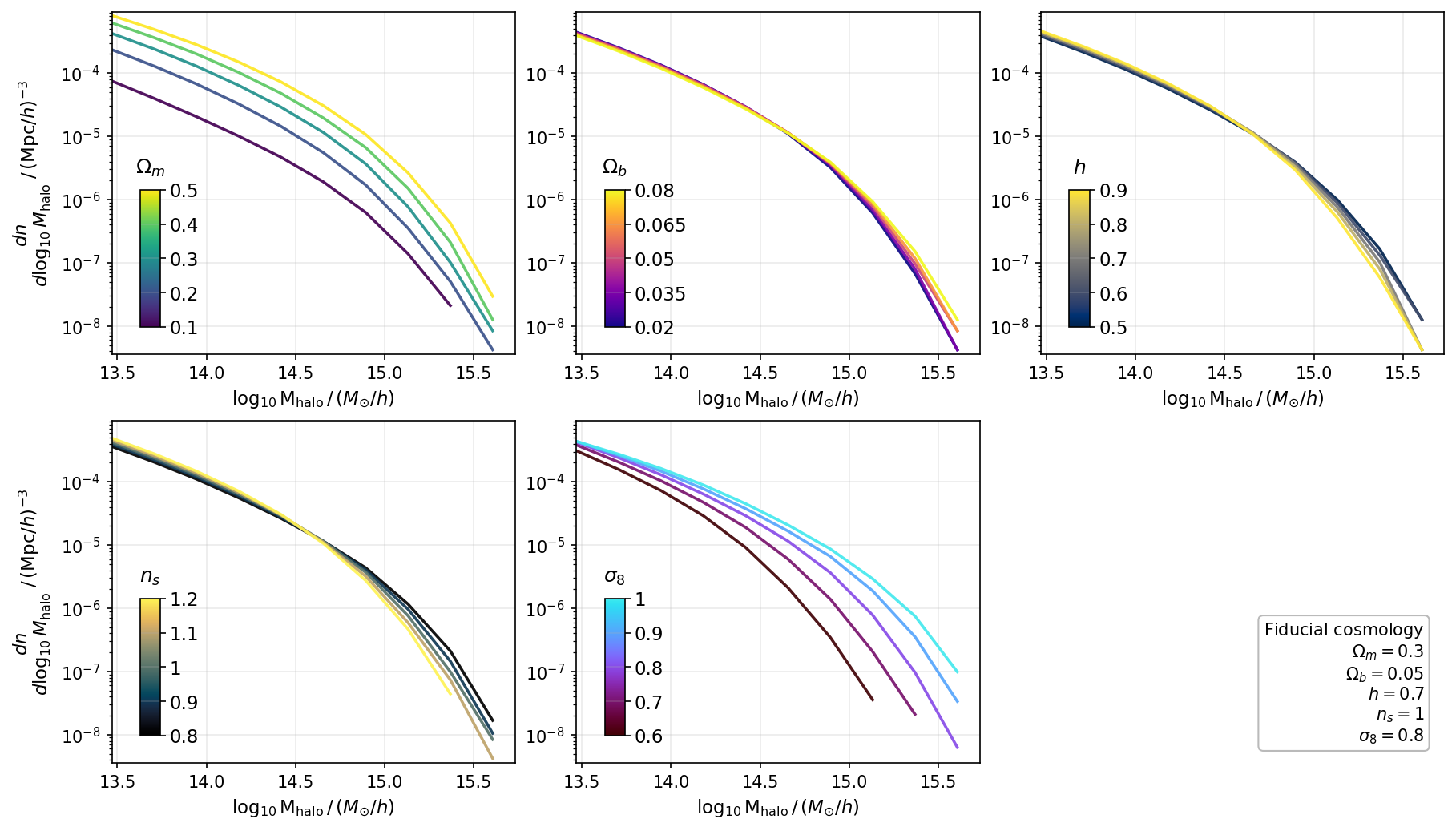}
    \caption{The predicted global HMF as a function of the cosmological parameters $\Omega_m$, $\Omega_b$, $h$, $\sigma_8$ and $n_s$. Each panel shows the impact of varying a single parameter whilst keeping the other parameters fixed to the fiducial values stated in the bottom right.    
    }
    \label{fig:global_hmf_cosmo}
\end{figure*}

In \fig{global_hmf_cosmo} we show how our predicted HMF varies with each cosmological parameter around some fiducial parameter set.
We can see the strong dependence of the normalisation on $\Omega_m$, and the sensitivity of the high mass tail to both $\sigma_8$ and $n_s$.

\subsection{How many regions are required, and what volume?}
\label{sec:how_many}

\begin{figure}
	\includegraphics[width=\columnwidth]{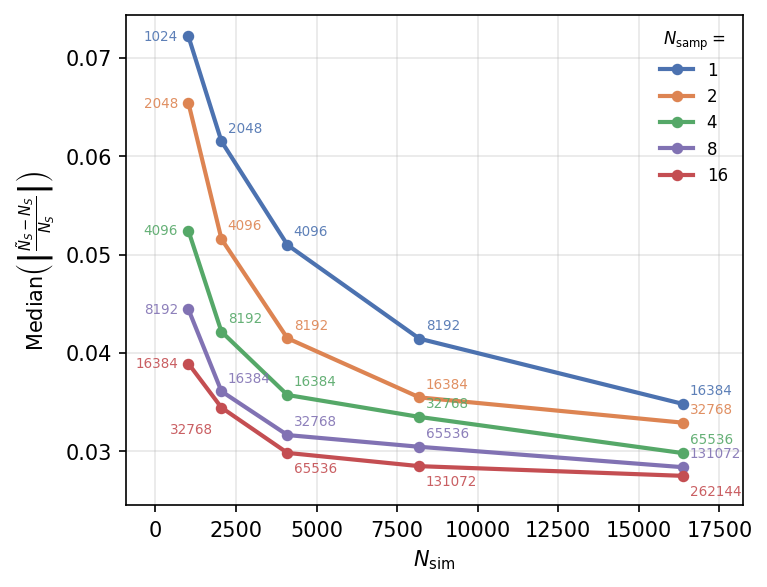}
    \includegraphics[width=\columnwidth]{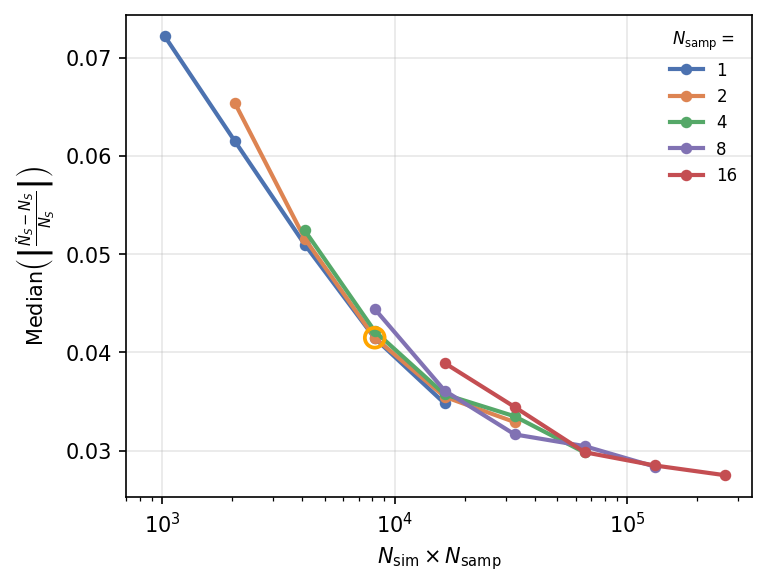}
    \caption{Median absolute fractional error across all mass bins as a function of (top panel) number of simulations $N_{\rm sim}$, and (bottom panel) total number of regions ($N_{\rm sim} \times N_{\rm samp}$). Coloured lines show the number of regions selected within each simulation. Numbers printed by each point in the top panel state the total number of regions used. The orange circle denotes our fiducial simulation and sample selection shown in Figures \ref{fig:conditional_hmf_examples}, \ref{fig:frac_error}, \ref{fig:global_hmf_examples} and \ref{fig:global_hmf_cosmo}.}
    \label{fig:region_sim_number}
\end{figure}

The purpose of this study has been to evaluate whether a suite of relatively small volume `zoom' simulations can be used to reconstruct global distribution functions, with the assumption that the HMF is one of the more challenging one-point distributions to emulate.
To this end, we now explore the sensitivity of our results to two key parameters: the number of training regions, and the volume of those regions.

The top panel of \fig{region_sim_number} shows the median absolute fractional error on the held out test set across all mass bins against the total number of simulations.
We also show separate lines for the number of regions sampled from each simulation, from 1 to 16.
In general, for a higher number of simulations the fractional error decreases. 
Similarly, for greater numbers of sample regions per simulation, the fractional error also decreases.
To better understand whether it is (1) the diversity in the overdensity sampling at fixed cosmology that is driving the error down, or (2) the diversity in cosmological parameters, we show the fractional error against total number of regions in the bottom panel of \fig{region_sim_number}. 
There is a strong correlation with total number of regions, but no preference for number of simulations or oversampled regions at fixed total number of regions.

\begin{figure}
    \includegraphics[width=\columnwidth]{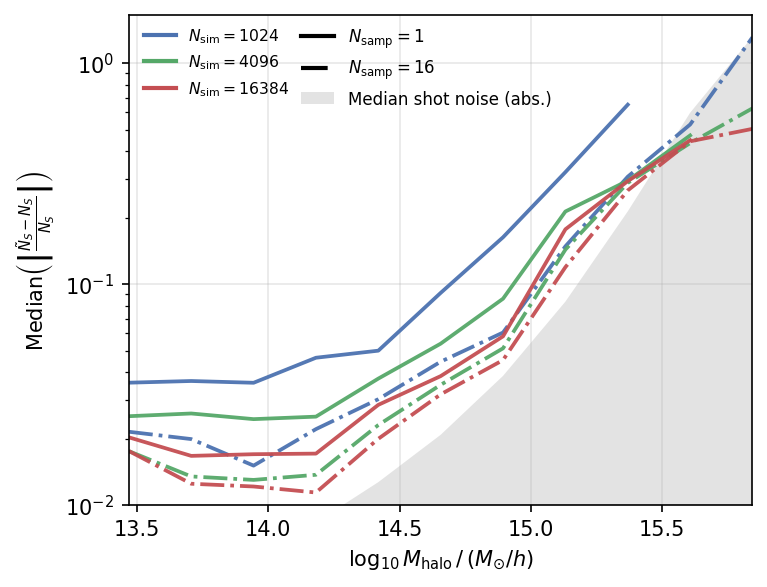}
    \caption{Median absolute fractional error against halo mass bin for all simulations in the test set.
    Coloured lines show different numbers of training simulations $N_{\rm sim}$, and solid and dash-dotted lines show either 1 or 16 samples from each simulation, respectively.
    The grey shaded region shows the median contribution from shot noise across the test set.}
    \label{fig:region_sim_mass}
\end{figure}

We also show the absolute fractional error in each mass bin in \fig{region_sim_mass}.
In general the absolute fractional error increases with mass, as expected due to the contribution of shot noise.
It reaches a floor at lower masses that is strongly dependent on the number of simulations and the number of regions selected, plateauing at around $\sim$4\% for 1024 simulations with a single region selected.
Increasing the number of regions selected per simulation to 16 reduces the low-mass bin error to $\sim$2\%; increasing the simulation number to 16384 whilst selecting just one region from each achieves similar errors.
In higher mass bins there is a similar decrease in the error toward the shot noise limit as the number of simulations or regions is increased.
The lowest errors are achieved with 16384 simulations and 16 regions from each simulation, a total of 262144 regions, reaching low-mass bin errors of $\sim$1\%.

\begin{figure}
	\includegraphics[width=\columnwidth]{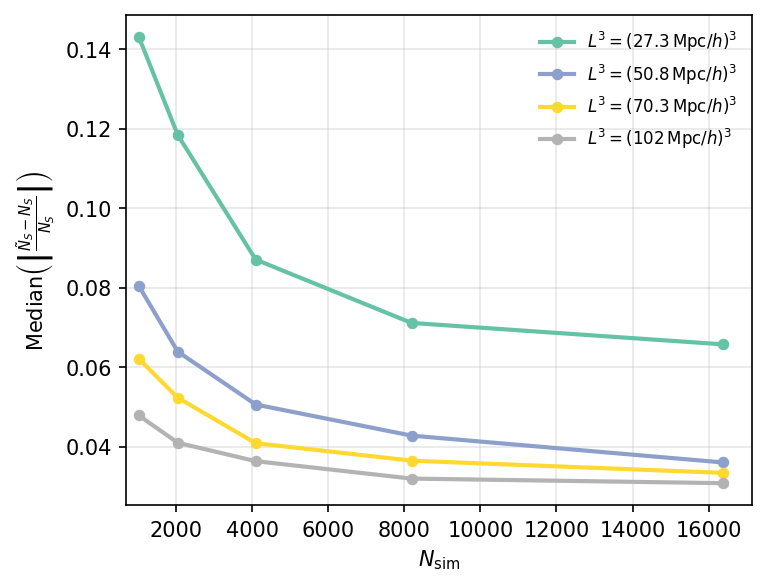}
	\includegraphics[width=\columnwidth]{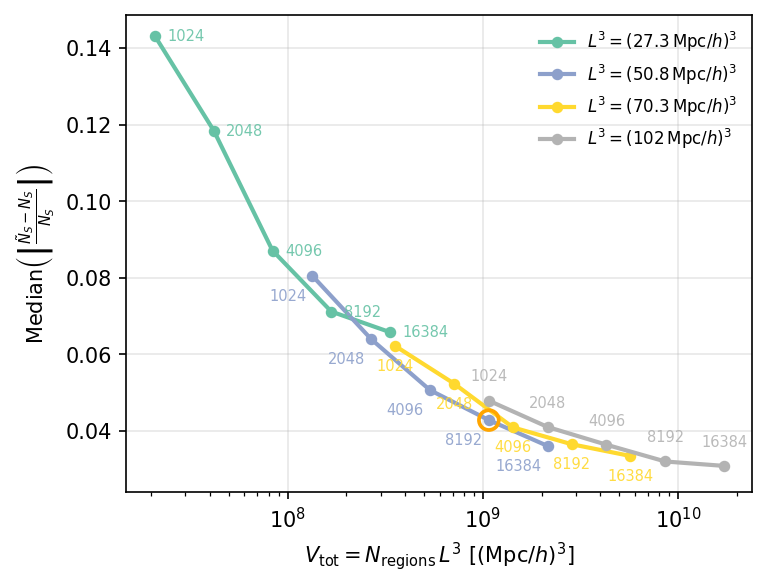}
    \caption{Impact of varying the volume of each selected region, from $(27.3 \; h^{-1}{\rm Mpc})^3$ to $(102 \; h^{-1}{\rm Mpc})^3$. \textit{Top}: Median absolute fractional error as a function of number of simulations across the test set (where a single region is selected from each simulation).
    \textit{Bottom}: Median absolute fractional error against total simulated volume selected across all regions.
    The orange circle denotes the fiducial sample size and region volume used in the rest of the analysis.}
    \label{fig:region_volume}
\end{figure}

\begin{figure}
    \includegraphics[width=\columnwidth]{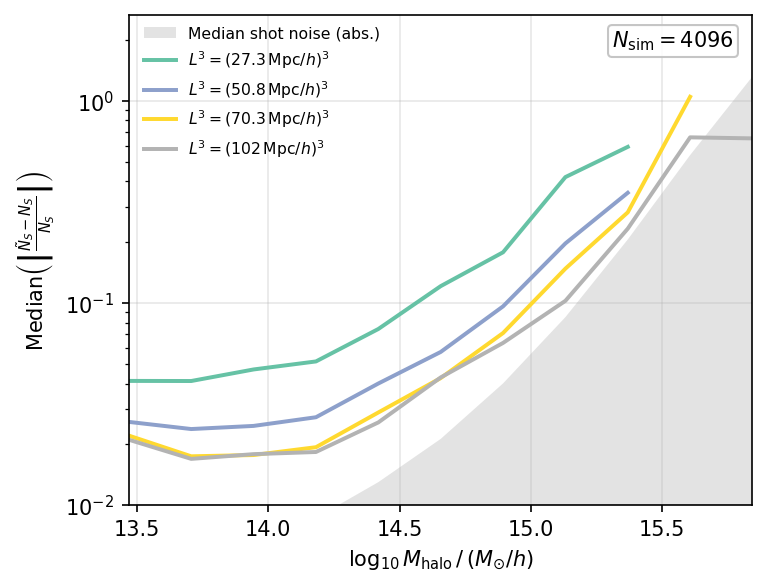}
    \caption{As for \fig{region_volume}, but showing median absolute fractional error against halo mass bin for all simulations in the test set, where we assume a single region selected from 4096 training simulations.
    The grey shaded region shows the median contribution from shot noise for all test set parent simulations.}
    \label{fig:region_volume_mass}
\end{figure}

\fig{region_volume} shows the impact of varying the volume of each region, for a fixed number of simulations and regions.
In general, for a fixed number of regions, larger regions lead to lower errors, since the total volume is larger. 
This can be seen explicitly in the bottom panel of \fig{region_volume}, where there is a strong trend in the absolute fractional error with total combined volume of all regions.
Larger regions also incur lower cosmic variance, making it easier for the model to learn the relationship between the input parameters and the HMF at fixed overdensity.
However, the improvement seems to plateau in regions around $(70 \, h^{-1}{\rm Mpc})^3$ in volume, with little improvement from going to $(100 \, h^{-1}{\rm Mpc})^3$.
For a fixed total volume, it seems that there is a minor improvement from having fewer larger regions, but this trend does not hold for the largest volumes.

Together, these results suggest that, using just 4096 regions selected from unique parent simulations, with volumes of $(50-70 \, h^{-1}{\rm Mpc})^3$ will enable training of an accurate HMF emulator with median absolute fractional error $\sim$5\% across all mass bins.
This is competitive with other emulators available to date trained on suites of periodic simulations \citep{mcclintockAemulusProjectII2019,bocquetMiraTitanUniverseIII2020a,saez-casaresEMANTISEmulatorFast2024,shenAemulus$n$Precision2025,chenCSSTCosmologicalEmulator2025}; in most cases we also provide predictions over a larger volume of cosmological parameter space.
It also far out performs traditional fitting functions \citep[e.g.][]{tinkerHaloMassFunction2008}, which have known errors of up to 30\% \citep[see][]{bocquetMiraTitanUniverseIII2020a}.

%% file: discussion.tex
\section{Discussion}
\label{sec:discussion}

\subsection{Zoom simulation suite applications}

The method we have presented will show the most promising applications when using zoom simulations of selected regions.
Zooms allow for increased resolution in the region of interest, as well as added physics, whilst preserving the influence of the field on large scales \citep{katzHierarchicalGalaxyFormation1993,tormenStructureDynamicalEvolution1997}.
The most common example is the inclusion of hydrodynamics and subgrid processes for galaxy evolution modelling \citep[e.g.][]{crainGalaxiesIntergalacticMediumInteraction2009,barnesClusterEAGLEProjectGlobal2017,hopkinsFIRE2SimulationsPhysics2017,lovellFirstLightReionization2021}.
By using a suite of zoom simulations with increased resolution and physical fidelity selected from low resolution parent dark-matter only simulations, the method will allow us to probe the halo mass function down to much lower halo masses than achievable in similar resolution periodic simulations, as well as other distribution functions including baryons, such as the stellar mass function.

We provide a quantitative illustrative example below.
If we treat the Quijote DMO simulations used in this study, with volume $(1 \; h^{-1}{\rm Gpc})^3$ and particle number 512$^3$, as our parent volumes, and selected regions with volume $(50 \; h^{-1}{\rm Mpc})^3$ from these to resimulate as zooms, each with $512^3$ particles, this would give a dark matter particle mass in each zoom (in median overdensity regions) of $m_p = \Omega_m \rho_{\rm crit} L^3 / N_p \approx 7.8 \times 10^{7} \; (\Omega_m / 0.3) \; h^{-1} \, {\rm M_\odot}$.
This is a factor of $\sim$8000 improvement over the parent boxes, allowing haloes of mass $\sim 10^{10} \, {\rm M_{\odot}}$ to be resolved, well within the characteristic halo occupation mass regime of emission line galaxies \citep{zhaiLinearBiasHalo2021,geachClusteringHaEmitters2012}.
Of course, each simulation is by construction \textit{not} necessarily median overdensity, reaching $\delta$ values of up to $\sim3$.
Through \eq{overdensity} this translates to a factor of 4 increase in the particle number, at fixed volume, or $\sim 812^3$ particles.
This is somewhat balanced by the reduced particle number at fixed volume for underdense regions; the most underdense regions ($\delta \sim -0.75$) would have $\sim 325^3$ particles.

% We can use these region particle numbers to estimate the overall computational cost of a suite of DMO zooms, using the total particle number of all regions combined as a proxy.
% For our fiducial tail-oversampled selection over 8192 regions, selected from 4092 parent simulations, this gives a total particle number of $\sim10\,300^3$, or a mean of $520^3$ particles per region.\footnote{The mean here is greater than the median due to the positively skewed overdensity distribution in linear overdensity space.}
A \textit{single} periodic $(1\, h^{-1}{\rm Gpc})^3$ volume simulation at the same resolution would require $10\,240^3$ particles, which would already place it within the Exascale regime \citep[e.g.][]{ishiyamaUchuuSimulationsData2021,frontiereCosmologicalHydrodynamicsExascale2025a,collaborationEuclidFlagshipGalaxy2025}.
For a suite of 4096 periodic simulations exploring the same breadth of cosmological parameters at the same resolution, this increases to $163\,840^3$ particles in total.
Ignoring the overheads associated with running zoom simulations, and any tail oversampling we therefore estimate that a zoom DMO suite can be run for approximately 0.025\% of the cost of an equivalent resolution periodic simulation suite.
Additionally, each of these $512^3$ simulations, including the most overdense, can fit on a single compute node on modern computing facilities.
Even those requiring multiple nodes will be much easier to schedule and run compared to a single monolithic periodic simulation.
We also assume here that the cost of the parent simulation from which the zooms are selected is significantly cheaper than each zoom; with modern Particle Mesh (PM) codes \citep[e.g.][]{liPmwdDifferentiableCosmological2022} this is the case at sufficient resolution to obtain accurate overdensities at $z = 0$ on $\gtrsim 50 \, h^{-1} {\rm Mpc}$ scales.

In summary, it is clearly feasible to run a suite of thousands of DMO zoom simulations with similar particle numbers to the Quijote BSQ suite, and possibly even full hydrodynamic suites,\footnote{We expect the number of required simulations to be considerably lower for a suite with a fixed cosmology.} from which accurate emulators over large dynamic ranges can be constructed.
The environmental dependence of galaxy evolution could also be explicitly explored in such a suite \citep[e.g.][]{lovellFirstLightReionization2021,simsCAMELSEnvironmentsImpact2026a}.

\subsection{Extensions to the model}
One factor we have not explored is extending the proposed framework to higher redshifts, by using the high redshift snapshots from each zoom simulation.
Since the selected Lagrangian region of each zoom simulation changes shape and volume over cosmic time, it is more challenging to use higher redshift snapshots from regions selected at lower redshift to measure volume normalised distribution functions, and train effective emulators (though for sufficiently large region volumes this volume change will be minimised).
One approach would be to approximate the volume of the zoom at higher redshift (using, for example, a Voronoi tesselation) and measure the relative overdensity by measuring the overdensity distribution on that scale from the low-resolution parent.
% For the most overdense regions the associated Lagrangian region in the initial conditions is much larger, and the Eulerian volume traced by this region at high redshift will be larger, so one could in principle subsample within that volume.
% For underdense regions the corresponding Eulerian volume traced becomes smaller at higher redshift, but this could be compensated by selecting a larger Lagrangian region, using a fixed particle number for all regions below the median overdensity.
We will explore these issues in future work.

Another interesting extension would be Active Learning (AL) to select the most informative regions in parameter space for new training regions, both in terms of cosmology and overdensity.
\cite{leeZoomingCARPoolGPLane2024} demonstrated a version of this AL approach called CARPoolGP, using a Gaussian Process emulator alongside a tiered simulation approach, to select new objects to improve the prediction accuracy of their emulator for group properties.
Unfortunately such an emulator is unsuitable for emulating the halo mass function, but similar approaches could be utilised to better inform the choice of training samples, and potentially significantly reduce the total number of simulations required.
It would also be interesting to evaluate whether the conclusions of \sec{how_many} still hold when applying AL, or whether new samples in cosmological parameter or overdensity space would perform better.

In this study we have focused on the HMF as an example of a distribution function which necessitates large volumes in order to probe the high mass end.
Extending to other scaling relations and distribution functions, such as the mass-concentration relation, will be trivial using a similar architecture \citep[e.g.][]{lovellHierarchyNormalizingFlows2023}, swapping out a more appropriate likelihood, or other galaxy-halo modelling techniques \citep[e.g.][]{nguyenHowDREAMSAre2026}.
Large volumes are also necessary for measuring higher order statistics that probe the scale-dependent clustering of haloes and galaxies.
Each selected region obviously does not allow the measuring of these statistics out to large scales (low wavenumber).
However, a number of techniques have been presented in recent years that provide an avenue for using these high fidelity simulations to probe the large scale clustering signal.
In \cite{lovellMachineLearningApproach2022} we trained an emulator for the galaxy-halo relationship on a combination of zooms and periodic simulations run with the EAGLE code \citep{schayeEAGLEProjectSimulating2015,barnesClusterEAGLEProjectGlobal2017}.
These could then be applied to large volume periodic DMO simulations, allowing us to predict the projected correlation function at much larger scales, and with both higher mass haloes (for which we had sufficient samples) and lower mass haloes (for which we had sufficient resolution).
Other compelling approaches include GOTHAM \citep{pandeyTeachingDarkMatter2024} and CHARM \citep{pandeyCreatingHalosAutoregressive2025}; the former utilises a transformer architecture, whereas the latter leverages stacks of spline flows, also trained on Quijote, to train an emulator for the halo spatial and property distribution, enabling the measurement of two-point and higher order statistics by generating field level predictions.
This method relies on features of the field measured on larger scales than the zoom \citep[see also][]{pirasFastRealisticLargescale2023,dingPineTreeGenerativeFast2024}; \cite{bartlettByebyeLocalinmatterdensityBias2024a} showed how using local-in-matter density (LIMD) features can bias predictions of higher order statistics, necessitating this expanded feature space.

\subsection{Parameter proposal distributions}
The range of each parameter varied in Quijote is relatively wide.
For example, $\sigma_8$ is varied between 0.6 and 1.0, but current qualitative uncertainties place $\sigma_8$ within 0.65 -- 0.9 across a range of probes.
This large parameter range significantly increases the parameter volume that must be modelled effectively; for $\sigma_8$ this is tightly linked to the maximum halo mass, requiring further expensive simulations in the high halo mass regime.
Reducing the $\sigma_8$ parameter range would not only reduce the number of high mass haloes required to cover the full dynamic range of masses in high $\sigma_8$ cosmologies, it would also allow for more simulations in the remaining volume of parameter space, permitting increased accuracy at fixed simulation cost. 

%% file: conc.tex
\section{Conclusions}\label{sec:conc}

We have demonstrated a new method for reproducing global distribution functions for arbitrary cosmological parameters using neural density estimators trained on small regions selected from the Quijote simulation suite.
Our findings are as follows:
\begin{itemize}
    \item We train a conditional emulator for the halo mass function (HMF), conditioned on the local overdensity percentile and five $\Lambda$CDM cosmological parameters ($\Omega_m$, $\Omega_b$, $h$, $n_s$, $\sigma_8$), using small regions ($L \sim 50 \, h^{-1}{\rm Mpc}$) selected from the Quijote BSQ simulation suite. By integrating over the overdensity distribution we recover the global HMF of the full $(1 \, h^{-1}{\rm Gpc})^3$ volume, using just $\sim 0.026\%$ of the original simulation volume.

    \item At the \textit{region} level, the emulator residuals are consistent with Poisson scatter across the full range of cosmologies, overdensity percentiles and halo mass bins in the test set. The signed deviance residuals show little systematic bias, and the predicted zero-count probabilities are well calibrated in all mass bins.

    \item The reconstructed \textit{global} HMF achieves a median fractional error of $\sim$2\% at low masses ($\lesssim 10^{14.5} \; {\rm M_\odot}$), converging toward the shot noise limit at higher masses. For 95\% of test cosmologies the fractional error remains below 5\% for halo masses below $10^{14.5} \; {\rm M_\odot}$.

    \item The emulator correctly captures the dependence of the HMF on cosmological parameters, including the strong sensitivity of the normalisation to $\Omega_m$ and of the high-mass tail to $\sigma_8$ and $n_s$.

    \item Emulator error is primarily dependent on the total volume simulated; sampling diversity in either cosmological parameters or range of environments does not lead to improved performance.
    Similarly, there is no strong preference between sampling many small volume regions over a few large regions, up to $\;\sim(70 \; h^{-1}{\rm Mpc})^3$, where there is a clear preference for more smaller-volume regions. Region volumes of $\;\gtrsim(50 \; h^{-1}{\rm Mpc})^3$ are sufficient to suppress cosmic variance to a level where further volume increases yield diminishing returns.
    With 4092 regions from unique parent simulations at this volume, median absolute fractional errors below 5.5\% are achievable.

    \item The model is built in JAX and is fully differentiable, enabling its direct use within gradient-based inference frameworks such as BlackJax. This makes it suitable for cosmological parameter estimation pipelines where the HMF is a key ingredient.
\end{itemize}

Our approach implies that suites of targeted zoom simulations, extracted from low resolution parent volumes with varying cosmological parameters, can emulate gigaparsec-scale distribution functions at a fraction of the cost of traditional periodic simulations.
This has implications for the interpretation of upcoming wide-field surveys on observatories such as \textit{Euclid}, \textit{Roman} and \textit{Rubin}.
By probing much lower halo masses than traditional periodic simulations, whilst still exploring large cosmological parameter ranges in Gigaparsec-scale volumes, the approach opens up the possibility of modelling tracers that inhabit much lower haloes over large scales, such as emission line galaxies.
By including full hydrodynamic models that capture baryonic effects, our method offers a practical route to producing the precise, parameter-dependent predictions these surveys demand, without the prohibitive cost of running large suites of Gigaparsec-scale hydrodynamic simulations.

%% file: appendix.tex
% \section{Empirical verification of the global integration method}

% In order to verify that the scheme proposed in section ?? for obtaining the global HMF works, we show some empirical tests.
% We use the true overdensity distribution, divided evenly into $N$ percentiles, and then find the nearest region with this overdensity in a given box. 
% We then apply the same integration scheme as shown in section ??, essentially summing each HMF and dividing by the total number of regions.
% This shows what the predicted global HMF would be \textit{given a perfect emulator}, with the caveat that there may still be some scatter at a given overdensity that will not be captured by this method. 